\documentclass[lettersize,journal]{IEEEtran}
\usepackage{amsmath,amsfonts}
\usepackage{algorithmic}
\usepackage{algorithm}
\usepackage{array}
\usepackage[caption=false,font=normalsize,labelfont=sf,textfont=sf]{subfig}
\usepackage{textcomp}
\usepackage{stfloats}
\usepackage{url}
\usepackage{verbatim}
\usepackage{lipsum} 
\usepackage{graphicx}
\usepackage{cite}
\usepackage{booktabs}
\usepackage{makecell}
\usepackage{enumitem}
\usepackage{graphicx} % For including images
\usepackage{hyperref} % For hyperlinks
\usepackage{tabularx} % For automatic column width adjustment

\begin{document}

\title{A Cyber-Resilient Learning-Driven Strategy for Smart Grids: An Att-BiLSTM–ConvGAN–MTD Hybrid Framework}

\author{
Ali~Peivand\,,
and Seyyed~Mostafa~Nosratabadi\,,~\IEEEmembership{Member, IEEE}}

% \thanks{Manuscript received ####, 2025.}}

% \markboth{IEEE Transactions on Emerging Topics in Computational Intelligence}%


\maketitle
\begin{abstract}
Wind power uncertainty and cyber-attacks, though stemming from different domains, their concurrent occurrence can critically heighten grid vulnerability. This paper presents a data-driven multi-stage hybrid deep learning framework for secure and cost-effective power system scheduling under wind uncertainty and cyber threats. The proposed model comprises three integrated modules: (i) an attention-enhanced bi-directional LSTM (Att-BiLSTM) network for short-term wind power forecasting, leveraging temporal and spatial features with Optuna-based hyperparameter tuning; (ii) a convolutional GAN (ConvGAN) module that generates high-fidelity wind scenarios to capture stochastic variability; and (iii) a cost-benefit optimization engine incorporating a Dynamic Defense Mechanism (DDM) that adaptively perturbs transmission line reactances to counter False Data Injection (FDI) attacks. Two configurations—with and without Battery Energy Storage (BES)—are examined on a modified Illinois 200-bus system to evaluate the economic and cybersecurity performance of the framework. Simulation results confirm the superiority of the proposed method, achieving 99\% and 98\% accuracy in wind forecasting and predictive scheduling, respectively. The integration of BES yields cost reductions exceeding 3.8\%, with full elimination of wind curtailment at maximum storage capacity. Furthermore, a detectability-cost trade-off analysis is conducted, evaluating optimal reactance perturbation levels and strategic line selection across four loading ranges. The findings highlight how BES utilization complements Moving Target Defense (MTD) to mitigate cost increments while enhancing cyber resilience.
\end{abstract}


\section*{Nomenclature}
\subsection{Acronyms}
\begin{description}[leftmargin=2.5cm, labelwidth=2cm, labelsep=0.5cm, align=left]
    \item[Att-BiLSTM]   Attention-enhanced Bi-directional Long Short-term Memory
    \item[BES]      Battery Energy Storage
    \item[CNN]      Convolutional Neural Network
    \item[ConvGAN]  Convolutional Generative Adversarial Network
    \item[DDM]      Dynamic Defense Mechanism
    \item[FDI]      False Data Injection
    \item[MTD]      Moving Target Defense
    \item[RFR]      Random Forest Regressor

\end{description}
\subsection{Indices}
\begin{itemize}[label={}, leftmargin=1.5cm, labelwidth=1cm, labelsep=0.5cm, align=left]
    \item[$t$] Time index
    \item[$bes$] Battery energy storage index
    \item[$gn$] Generator index
    \item[$wf$] Wind farm index
    \item[$d$] Discriminator component index of the GAN model
    \item[$g$] Generator component index of the GAN model
    \item[$l$] Neural network layers
    \item[$ob$] Observation index in dataset
    \item[$br$] Network's lines
    \item[$ar$] Attack index in DDM model
\end{itemize}
\subsection{Parameters}
\begin{itemize}[label={}, leftmargin=2.3cm, labelwidth=1.8cm, labelsep=0.5cm, align=left]
    \item[$\Pr_{ng}^{su}, \Pr_{ng}^{sd}$] Start-up and shut-down prices for ${ng}^{th}$ generator
    \item[$S_{bes}^{ch}, S_{bes}^{dis}$] Charge and discharge price of battery energy storage
    \item[$\mathit{H}_c^{ar}$] Projection of the $ar^{th}$ attack vector onto perturbed subspace
    \item[$\eta_{bes,ch}$, $\eta_{bes,dis}$] BES charging and discharging efficiencies
    \item[$A$, $B$] Incidence and branch-to-bus matrics
    \item[$c$] Attack parameter vector
    \item[$P_D(\mathit{a})$] Probability of detecting attack $\mathit{a}$ under MTD
    \item[$\mathit{A}'(\delta)$] Set of detectable attacks with probability $> \delta$
    \item[$\lambda(\cdot)$] Lebesgue measure used to quantify volume
    \item[$\eta'(\delta)$] Normalized detectability ratio of $\mathit{A}'(\delta)$ to $\mathit{A}$
    \item[$A_1, A_2, A_3$] Cost coefficients of generators
    \item[$H_t, H_t^{'}$] Measurement matrics before and after changing line reactances at time $t$
    \item[$\lambda_{wf}$] Wind power penalty at time $t$
    \item[$\Lambda$] Total number of simulated attack
    \item [$\boldsymbol{\chi}_t$, $\boldsymbol{\zeta}_t$] Hidden state output and Input vector in the LSTM unit
    \item [$\boldsymbol{\xi_{\max}}$] Maximum allowed norm of attack vector
    \item [$\delta$] Predefined detection probability threshold
    \item [$\mathcal{A}$] Full set of attack vectors with bounded norm
    \item [$\mathcal{A}^{'}$] Subset of feasible attack vector space
    \item [$N_g, N_b, N_{ob}$] Total number of generators, batteries, and wind data observations

\end{itemize}
\subsection{Variables}
\begin{itemize}[label={}, leftmargin=2.7cm, labelwidth=2.2cm, labelsep=0.5cm, align=left]
    \item[$p_{gn}^t$] Scheduled power of ${gn}^{th}$ generator at time $t$
    \item[$P_{bes, ch}^t$, $P_{bes, dis}^t$] Charging and discharging power of energy storage
    \item[$P_{wf}^t$] Wind farm output at time $t$
    \item[$E_{bes}^t$] Energy level of energy storage at time $t$
    \item[$\alpha_{gn}^t$, $\beta_{gn}^t$, $u_{gn}^t$] On, off, and current states of ${gn}^{th}$ generator at time $t$
    \item[$z_{bes, ch}^t, z_{bes, dis}^t$] Charging / discharging state of ${bes}^{th}$ BES at time $t$
    \item[$D_t$] Diagonal matrix of the reciprocals of branch reactances
    \item[$x_{br}$] Reactance of $br^{th}$ line
    \item[$C_0, C_{DDM}$] OPF-related cost before and after applying DDM 
\end{itemize}

\section{Introduction}
\IEEEPARstart{T}{he} modern power system has undergone significant transformations in recent years, driven by the dual need to minimize operating costs and maximize reliability. However, the growing integration of renewable energy sources—particularly wind, with its inherent uncertainty—and the rise in cyber threats due to accelerated digitalization present significant challenges to both the economic and security dimensions of power system operation. While the levelized cost of wind energy has dropped by over 35\% between 2010 and 2021 \cite{eladl2024}, encouraging more investments in wind farms, this growth comes with operational risks. Notably, the variability in wind generation threatens grid reliability, increases reserve requirements, and contributes to wind curtailment. For example, Spain curtailed over 1100 GWh of wind energy in 2013, while China's wind curtailment dropped significantly from 17\% in 2016 to 3\% by 2020 following improved planning mechanisms \cite{chen2022}. At the same time, the increasing complexity and frequency of cyberattacks on critical infrastructure further complicate the secure and economic operation of modern power systems \cite{CC}.

To address these two critical issues—achieving cost-effective scheduling and ensuring cybersecurity resilience—recent advancements in Artificial Intelligence (AI), particularly in deep learning, have shown promise. For economic optimization, accurate forecasting of wind power is essential to mitigate intermittency and reduce curtailment. For cybersecurity, strategies like the Moving Target Defense (MTD) have been introduced, which proactively alter system parameters (e.g., transmission line reactance) to invalidate the attacker’s knowledge of the system \cite{10854990}. However, most existing MTD-based approaches either overlook the economic implications of frequent system perturbations or fail to incorporate them within a broader optimization framework. Moreover, traditional forecasting methods such as Auto-Regressive Integrated Moving Average (ARIMA) or Seasonal-ARIMA \cite{zhang2022} fall short when handling high-dimensional, nonlinear data inherent to renewable generation. Thus, AI-enabled techniques—particularly deep neural networks—offer a more capable alternative, both for wind power forecasting and for enhancing the reliability of secure scheduling under cyber threats. Several studies have confirmed that improved forecasts can reduce operating costs and curtailment \cite{LSTM1,peivand2024}, while also supporting resilient grid operation under attack scenarios \cite{dkhili2023}.

Despite growing interest in both forecasting and cyber-physical security of power systems, most existing studies have addressed these components in isolation. Some works focus solely on improving renewable energy forecasting through deep learning models \cite{Cui2023}, while others design secure control mechanisms to detect and mitigate FDI attacks \cite{ESS2}. However, few studies have attempted to unify these domains into an integrated scheduling pipeline. More importantly, existing MTD-based strategies often neglect the operational and financial impacts of line perturbations, failing to provide a realistic cost-benefit analysis. Additionally, conventional scenario generation techniques rely on statistical assumptions or Monte Carlo methods, which may not capture the complex temporal patterns of renewable variability \cite{tran2021}.

Another critical yet often overlooked component in the secure and economic operation of modern power systems is the role of Battery Energy Storage Systems (BES) combined with ANN \cite{liu2024}. These systems not only enable better management of renewable intermittency—by shifting surplus wind generation to periods of high demand—but also improve the grid’s flexibility and resilience to unexpected events, including cyber disruptions. As proven by authors in  \cite{li2023}, utilizing the CNN-LSTM method, investigate state-of-charge (SOC) estimation which can be crucial for reliability. Also, In \cite{memarzadeh2023}, a forecasting LSTM-designed model empowered by genetic algorithm is used to tackle with renewable energy uncertainty and energy storage sizing. In \cite{liu2023}, charging and discharging optimal scheduling are determined by a deep Q-network structure. Prediction and optimization of energy storage along with other power system participants ensure maintaining reliable and stable operation of power system. Despite their growing deployment worldwide, many existing studies treat storage as a static component or exclude its strategic integration in optimization under adversarial scenarios. Our work fills this gap by evaluating both cases—with and without BES—to highlight how storage can mitigate wind power curtailment, reduce operational costs, and enhance cyber-resilience through more flexible scheduling options \cite{BES}.

To overcome these limitations, a learning-augmented, multi-stage hybrid deep framework is introduced. This framework is composed of three integrated modules. The first module employs an Attention-based Bidirectional LSTM (Att-BiLSTM) to perform precise short-term wind power forecasting by leveraging a variety of temporal features; (ii) a Convolutional GAN (ConvGAN) model that generates synthetic wind scenarios based on the predicted data, capturing temporal dependencies and data-driven stochasticity; and (iii) a cost-benefit optimization engine equipped with a Dynamic Defense Mechanism (DDM) that perturbs transmission line reactances in a selective and economical way to thwart stealthy False Data Injection (FDI) attacks. By dynamically modifying network parameters, DDM obscures the predictability of the system’s operational state, thereby complicating the efforts of potential cyber attackers and reinforcing the overall security framework \cite{ghiasi2023}. For instance, authors in \cite{lakshminarayana2021} employed a Moving-Target Defense model which aims to constantly alter the lines’ reactances, distorting attackers’ information about the system. The symbiotic relationship between DDM and BES transcends conventional benefits, serving a dual purpose: it substantively bolsters the physical robustness of the power infrastructure and shields it against cyber threats. This dual strategy not only mitigates operational costs and reduces instances of power curtailment but also enhances the grid's ability to withstand and adapt to disruptions, thereby elevating its overall resilience. 
The combination of these technologies ensures a fortified defense against both physical failures and cyber intrusions, establishing a more reliable and secure grid environment. Despite numerous studies investigating BES and DNN's impact on power system efficiency, some critical research gaps remain as: 1) a scarcity of studies bridging cost-efficient and cybersecurity-based optimization frameworks, particularly those leveraging a deep-based model incorporating Att-BiLSTM for forecasting and ConvGAN for scenario generation, 2) limited investigation of BES impacts within economic and cybersecurity optimization frameworks, and 3) insufficient predictive analysis aimed at reducing wind curtailment while simultaneously enhancing cybersecurity. Addressing these challenges, our model proposes a dual strategy integrating advanced predictive analytics with robust cybersecurity measures to not only enhance grid stability but also reduce vulnerability to cyber-attacks. Briefly, the main contributions of the proposed model are as follows:
\begin{itemize}
    \item Development of a multi-stage deep-based framework that properly integrates wind power forecasting (Attention-based Bidirectional-LSTM), scenario generation (ConvGAN), and secure economic scheduling.
    
    \item Application of Convolutional GAN (ConvGAN) to generate highly realistic, stochastic wind power scenarios that preserve realistic temporal patterns.

    \item Explore the trade-off between detectability and cost, seeking optimal perturbation levels and target line groups based on loading categories, and highlighting the existence of Pareto-optimal defense configurations. More, two comparative case studies—with and without BES—demonstrating that BES integration mitigates curtailment and offsets the cost impact of MTD deployment.
\end{itemize}
The remainder of this paper is organized as follows: The methodology section outlines the overall optimization procedure of the proposed model in section II. The third section presents the results of the proposed framework. Finally, the conclusion is provided in the last section.

\section{Methodology}
\begin{figure*}[!t]
\centering
\includegraphics[width=7in]{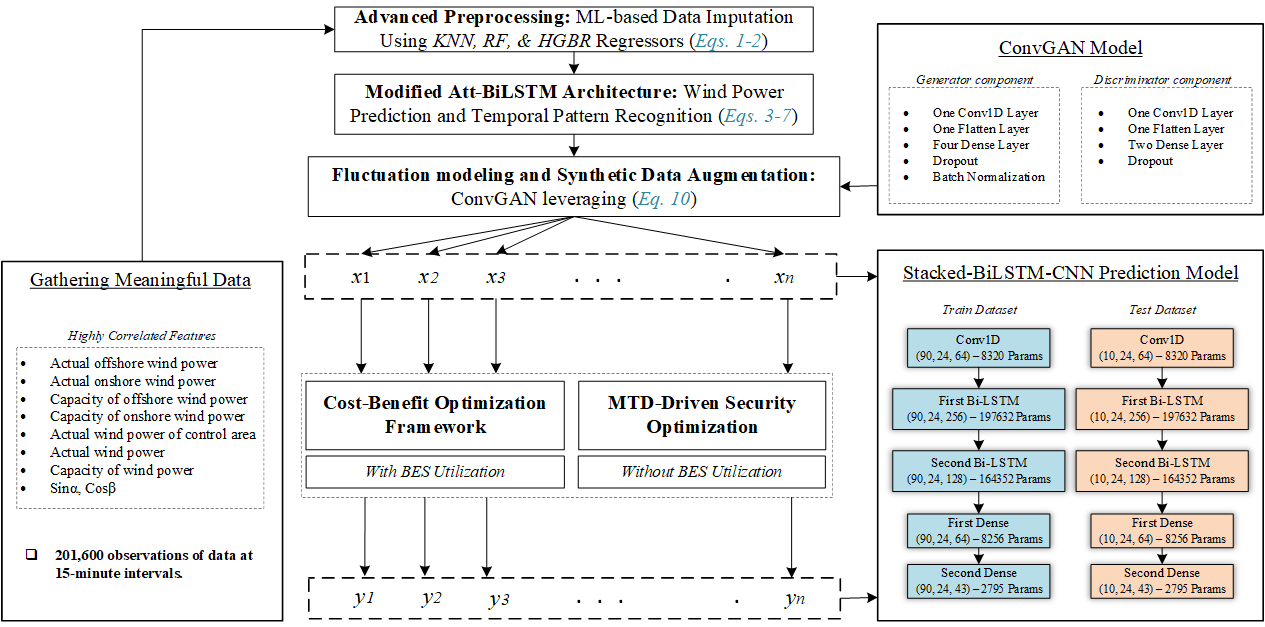}
\captionsetup{justification=centering}
\caption{Detailed workflow of the proposed model: integrating cost-efficient and security-based optimization framework.}
\label{fig_1}
\end{figure*}

The overall procedure of the proposed model is demonstrated in this section. As mentioned, this paper proposes a hierarchy multi-stage framework to predict the optimal scheduling of a power system including BES. The overall detailed workflow of the proposed model is demonstrated in Fig. \ref{fig_1}. The workflow begins with gathering essential data, including features such as wind power and capacity, which undergo advanced preprocessing using machine learning-based imputation techniques to handle missing values. Following this, an enhanced Att-BiLSTM model predicts short-term wind power by capturing temporal patterns, enhancing prediction accuracy. ConvGAN is then applied to model fluctuations and generate synthetic scenarios, simulating diverse wind power generation conditions. The proposed scheduling model is formulated as a Mixed-Integer Nonlinear Programming (MINLP) problem. Power system optimization is implemented using the \texttt{Pandapower} and \texttt{PYomo} libraries, which supports AC-OPF modeling and employs the IPOPT solver to handle nonlinearities, utilizing warm-start techniques for improved convergence. \cite{learn2001}.  

Finally, a stacked BiLSTM-CNN model evaluates the predictive performance on training and testing datasets, demonstrating the effectiveness of this integrated deep learning and optimization approach in reducing wind curtailment, improving economic efficiency, and guaranteeing the robust maintenance of power system under topology-aware cyber attacks.

\subsection{Data Recovery via Multi-Regressor Imputation}
During wind forecasting, missing data can significantly deteriorate model accuracy and generalization. To address this, a robust imputation framework is employed based on three ML regressors—Random Forest Regression (RFR), K-Nearest Neighbors (KNN), and Histogram-Based Gradient Boosting Regression (HGBR)—to reconstruct incomplete records before wind power prediction. Let $\mathcal{D} = {(\mathbf{x}_i, y_i)}$ and $i \in \{1, ..., N\}$ denote the dataset, where $\mathbf{x}_i \in \mathbb{R}^d$ is the input feature vector and $y_i \in \mathbb{R}$ is the target (e.g., wind power). The set $\mathcal{D}_{\text{miss}} \subset \mathcal{D}$ contains entries with missing $y_i$, for which we estimate $\hat{f}(\mathbf{x}_i) \approx y_i$.

\vspace{0.5em} \noindent \textit{Random Forest Regression:} RFR builds an ensemble of $K$ randomized decision trees ${h_k}$ and $k \in\{{1, ..., K\}}$, each trained on a bootstrapped subset. The final prediction is: \begin{equation} \hat{f}_{RFR}(\mathbf{x}) = \frac{1}{K} \sum_{k=1}^{K} h_k(\mathbf{x}). 
\end{equation}
The model’s robustness is evaluated via the margin function $mg(X, Y) = \mathbb{E}_k[I[h_k(X) = Y]] - \max_{j \neq Y} \mathbb{E}_k[I[h_k(X) = j]]$ and its generalization error is defined as $PE^* = \mathbb{P}[{X,Y}[mg(X,Y) < 0]$, where X and Y are the distribution of random vectors \cite{learn2001}.

\vspace{0.5em} \noindent \textit{K-Nearest Neighbors:} KNN estimates the missing value $y^*$ using a weighted average over $K$ nearest neighbors: \begin{equation} \hat{f}_{KNN}(\mathbf{x}) = \frac{1} {\sum_{i\in \mathcal{N}_{K}(\mathbf{x})} w_i} \sum_{i \in \mathcal{N}_{K}(\mathbf{x})} w_i \cdot y_i 
\end{equation} 
where $w_i = 1/|\mathbf{x} - \mathbf{x}_i|^2$ and $\mathcal{N}_K(\mathbf{x})$ denotes the $K$ closest points in Euclidean space.

\vspace{0.5em} \noindent \textit{Histogram-Based Gradient Boosting Regression:} The last regressor, HGBR, incrementally fits additive regression trees $T_m$ to the negative gradient of the loss function: $\hat{f}_{HGBR}(\mathbf{x}) = {\sum_{m=1}^{M}} \gamma_m T_m(\mathbf{x})$ where $\gamma_m$ is the learning rate. Histogram binning reduces memory usage and accelerates training on large datasets \cite{Joseph2024}.
\subsection{Wind Power Prediction \& Performance Evaluation}
To effectively capture temporal dependencies in wind data, the framework employs an enhanced Att-BiLSTM model, utilizing an attention mechanism. The proposed framework leverages forward and backward LSTM layers to extract sequential patterns and employs a soft alignment strategy to selectively focus on relevant time steps during the prediction phase \cite{DADA}. The mathematical formulation of the Att-BiLSTM model is presented in equations~\eqref{eq3}–\eqref{eq7}. At each time step $t$, the input vector $\boldsymbol{\zeta}_t$ is processed through the standard LSTM gates: input ($\mathbf{i}_t$), forget ($\mathbf{f}_t$), and output ($\mathbf{o}_t$), each computed via learnable parameters $\mathbf{W}_j, \mathbf{U}_j$, and bias $\boldsymbol{\theta}_j$:
\begin{flalign}
\mathbf{g}_t^{(j)} = \phi_j(\mathbf{W}_j \boldsymbol{\zeta}_t + \mathbf{U}_j \boldsymbol{\chi}_{t-1} + \boldsymbol{\theta}_j),\quad j \in \{\text{i, f, o}, \tilde{c} \}
\label{eq3}
\end{flalign}
Here, $\phi_j$ denotes the activation function: $\sigma(\cdot)$ for input, forget, and output gates, and $\tanh(\cdot)$ for the candidate cell state $\tilde{\mathbf{c}}_t$. The cell state is updated by blending the previous state and the new candidate is calculated by \eqref{eq4}:
\begin{flalign}
\mathbf{c}_t = \mathbf{f}_t \odot \mathbf{c}_{t-1} + \mathbf{i}_t \odot \tilde{\mathbf{c}}_t
\label{eq4}
\end{flalign}
The hidden state output $\boldsymbol{\chi}_t$ is computed as $\boldsymbol{\chi}_t = \mathbf{o}_t \odot \tanh(\mathbf{c}_t)$, and subsequently, attention weights are derived by aligning each hidden state with a learnable context vector $\mathbf{v}_a$ to capture temporal relevance.

\begin{flalign}
e_t = \mathbf{v}_a^\top \tanh(\mathbf{W}_a \boldsymbol{\chi}_t + \boldsymbol{\theta}_a)
\label{eq5}
\end{flalign}
These raw scores are normalized via the softmax function which is shown in~\eqref{eq6}.
\begin{flalign}
\alpha_t = \frac{\exp(e_t)}{\sum_{k=1}^{T} \exp(e_k)}
\label{eq6}
\end{flalign}
The final context vector is a weighted sum over all hidden states as shown in \eqref{eq7}
\begin{flalign}
\mathbf{c} = \sum_{t=1}^T \alpha_t \boldsymbol{\chi}_t
\label{eq7}
\end{flalign}
The term bi-directional in our proposed Att-BiLSTM structure, indicates the hidden representation at each time step which is a concatenation of forward and backward LSTM outputs as $\boldsymbol{\chi}_t = [\overrightarrow{\boldsymbol{\chi}_t};\ \overleftarrow{\boldsymbol{\chi}_t}]$. In that,  {$\overrightarrow{\boldsymbol{\chi}_t}$} points forward and {$\overleftarrow{\boldsymbol{\chi}_t}$} points backward. Moreover,  
the OPTUNA optimizer \cite{akiba2019} is employed here to adjust the hyperparameters of DNN's model. In terms of performance assessment, some prevalent evaluation metrics have been used which are defined as follows:
\begin{equation}
\label{eq8}
\text{MSE} = \frac{1}{N_{ob}} \sum_{{ob}=1}^{N_{ob}} (y_{ob}^t - y_{ob}^p)^2 \tag{8}
\end{equation}
where $y_{ob}^t$ and $y_{ob}^p$ indicate the target and prediction values of $ob^{th}$ observation in the mentioned data frame. Moreover, $N_{ob}$ refers to the total number of observations which is 201600 samples.
\subsection{ConvGAN Model}
To generate a synthetic data set that is highly similar to the original data set, this paper outlines a convolutional-based structure for the GAN model, ConvGAN for short, which is practicable in learning complex patterns and generating synthetic data with the highest similarities. The objective function of ConvGAN model is defined in~\eqref{eq9}. As the GAN model contains two opponent components, the generator, and discriminator, its objective function embodies two terms, representing generator and discriminator error values \cite{GAN1}.
\begin{equation}
\label{eq9}
\begin{split}
\max_{d} \min_{g} \mathcal{O}(d, g) &= \underbrace{\mathbb{E}_{x \sim p_{\text{data}}} [\log d(x)]}_{\text {Real sample}} \\
&\quad + \underbrace{\mathbb{E}_{z \sim p_{g}} [\log(1 - d(g(z)))]}_{\text {Generated sample rejection}}
\end{split}
\tag{9}
\end{equation}
In the proposed ConvGAN framework, the generator network receives noise vectors drawn from a uniform distribution and passes them through a stack of Conv1D, Dense, BatchNormalization, and Dropout layers to generate synthetic wind power profiles over a 24-hour horizon. The generator employs a dropout rate of 30\%, while the discriminator uses a similar convolutional architecture to distinguish real wind profiles from synthetic ones. Both networks are trained using the Binary Cross-Entropy (BCE) loss function and optimized using the Adam optimizer \cite{kingma2017adammethodstochasticoptimization}.
\subsection{Scheduling Model}
This section outlines the mathematical formulation of the power system scheduling model, including the objective function and associated constraints. All previously obtained outputs—such as wind power forecasts and generated scenarios—are integrated into the system optimization framework, which is performed under two configurations: with and without BESs. Additionally, the proposed framework is primarily developed from the perspective of an aggregator that owns and operates both conventional generation units and wind farms, aiming to maximize overall profitability while enhancing system reliability and cybersecurity. In this context, the aggregator acts as a centralized decision-making entity that leverages a unified forecasting-scheduling pipeline to optimize internal operations. The objective function, formulated in~\eqref{eq10}, consists of four main components. The first and second terms represent the startup and shut-down costs of conventional generators, along with their fuel consumption costs. The third term accounts for the operational cost associated with wind curtailment, while the last term incorporates degradation-related costs of BES systems.
\begin{align}
\label{eq10}
\min_{\substack{
\mathbf \Gamma
}}
\sum_{t=1}^{T} \Big(
& {\boldsymbol{\alpha}_{gn}^{t}}^\top \mathbf{Pr}_{gn}^{su} + {\boldsymbol{\beta}_{gn}^t}^\top \mathbf{Pr}_{gn}^{sd} \nonumber \\
& + (\mathbf{P}_{gn}^{t})^\top \mathbf{A_1} \mathbf{P}_{gn}^{t} + \mathbf{A_2}^\top \mathbf{P}_{gn}^{t} + \mathbf{1}^\top \mathbf{A_3} \nonumber \\
& + (\mathbf{z}_{bes,ch}^t \odot \mathbf{S}_{bes}^{ch})^\top \mathbf{P}_{bes,ch}^{t} \nonumber \\
& + (\mathbf{z}_{bes,dis}^t \odot \mathbf{S}_{bes}^{dis})^\top \mathbf{P}_{bes,dis}^{t} \nonumber \\
& + \boldsymbol{\lambda}_{wf}^\top \mathbf{P}_{wf}^t
\Big)
\tag{10}
\end{align}
Here, the decision vector is defined as $\boldsymbol{\Gamma} \in \mathbb{R}^{n}$, where $\boldsymbol{\Gamma} = \{ \mathbf{P}_{gn}^t, \boldsymbol{\alpha}_{gn}^t, \boldsymbol{\beta}_{gn}^t, \mathbf{P}_{bes, ch}^t, \mathbf{P}_{bes, dis}^t, \mathbf{P}_{wf}^t \}$ aggregates all time-dependent variables across $t \in \{1,\dots,T\}$. Additionally, $\mathbf{P}_{gn}^{t} \in \mathbb{R}^{N_g}$ denotes the real power output of $N_g$ conventional generators at time slot $t$. The startup and shut-down binary vectors are $\boldsymbol{\alpha}_{gn}^t, \boldsymbol{\beta}_{gn}^t \in \{0,1\}^{N_g}$, and their associated cost coefficients are $\mathbf{Pr}_{gn}^{sd}, \mathbf{Pr}_{gn}^{su} \in \mathbb{R}^{N_g}$. The $\lambda_{wf}$ represents the cost coefficient of each megawatt of wind power generated, based on $\$/MW$, and its value is 0.0001. The degradation cost of BES is captured using $\mathbf{P}_{bes, ch}^t, \mathbf{P}_{bes, dis}^t \in \mathbb{R}^{N_B}$, representing charging and discharging power vectors at time $t$, and their associated cost coefficients $\mathbf{S}_{bes}^{ch}, \mathbf{S}_{bes}^{dis} \in \mathbb{R}^{N_B}$.

\subsection{Power Systems Case Studies}
We consider two case studies to implement the proposed model. In the first case, the power system optimization is carried out without considering the BESs, focusing solely on the generators and wind farm. The main purpose of this case is to analyze the base operating cost of the system without considering the BESs. Hence, the BES-related terms are excluded in this case. To ensure accurate system representation, the AC power flow constraints are enforced at each bus $\iota \in \mathcal{B}$. These equations govern the steady-state balance of power flows in the network and are defined as follows:
\begin{equation} P_\iota^{\text{inj}} = \sum_{\kappa \in \mathcal{B}} |U_\iota||U_\kappa| \left( G_{\iota\kappa} \cos\varphi_{\iota\kappa} + B_{\iota\kappa} \sin\varphi_{\iota\kappa} \right) \tag{11} \end{equation} \begin{equation} Q_\iota^{\text{inj}} = \sum_{\kappa \in \mathcal{B}} |U_\iota||U_\kappa| \left( G_{\iota\kappa} \sin\varphi_{\iota\kappa} - B_{\iota\kappa} \cos\varphi_{\iota\kappa} \right) \tag{12} \end{equation}
Here, $P_\iota^{\text{inj}}$ and $Q_\iota^{\text{inj}}$ denote the active and reactive power injections at bus $\iota$, while $U_\iota$ and $U_\kappa$ are voltage magnitudes at buses $\iota$ and $\kappa$. The angle difference $\varphi_{\iota\kappa} = \varphi_\iota - \varphi_\kappa$ captures the phase shift, and $G_{\iota\kappa}$, $B_{\iota\kappa}$ represent the conductance and susceptance between buses $\iota$ and $\kappa$, respectively. In the second case, the optimal scheduling of the power system is determined with BES integration, which is expected to contribute to a reduction in both costs and wind curtailment. Three units of BES are utilized in this study, installed at buses 10, 20, and 30 of llinois 200-bus system with a maximum capacity of 250 MW \cite{birchfield2016}. The energy level of the BESs can be calculated using~\eqref{eq13}, which accounts for the charging and discharging efficiencies, denoted by $\eta_{bes, ch}$ and $\eta_{bes, dis}$, respectively. These efficiencies are set to 0.95 \cite{ESS1}.
\begin{equation}
\label{eq13}
E_{bes}^t = E_{bes}^{t-1} + \Delta t\left( \eta_{ch}^{bes} P_{bes,ch}^{t} - \frac{P_{bes, dis}^{t}}{\eta_{dis}^{bes}} \right)
\tag{13}
\end{equation}
Furthermore, to ensure cyclic energy balance, we impose a complementary constraint on the BESs, i.e., $E_{\text{bes}}^{\text{T-initial}} = E_{\text{bes}}^{\text{T-final}}$ which prevents the system from artificially benefiting by overcharging or undercharging the storage at the end of the scheduling horizon. Additionally, to ensure the feasible operation of the BESs, two mandatory constraints, defined by \eqref{eq14} and~\eqref{eq15}, are imposed. These constraints guarantee that the storage operates within its capacity and power limits.
\begin{equation}
\label{eq14}
0 \leq P_{bes,ch}^t \leq z_{bes,ch}^tP_{bes}^{\max} \tag{14}
\end{equation}
\begin{equation}
\label{eq15}
0 \leq P_{bes,dis}^t \leq z_{bes,dis}^tP_{bes}^{\max} \tag{15}
\end{equation}
The highest power capacity for a single battery is 250 MW, and with three batteries in the system, the total capacity reaches 750 MW. The charging and discharging costs are set to be 2.5 \$ per MWh, respectively. These constraints ensure that the BESs operate safely and efficiently while adhering to system limitations.

\subsection{Quantifying the Effectiveness of MTD via Detectable Attack Volume}
To assess the efficacy of the proposed MTD strategy in enhancing attack detection, a volume-based metric is employed. Let ${H}_t$ denote the system measurement matrix prior to reactance perturbation, constructed as:
\begin{align}
\label{eq16}
{H}_t = \begin{bmatrix}
D_t A^\top \\
- D_t A^\top \\
A D_t A^\top
\end{bmatrix}, \quad
D_t = \text{diag}\left(\frac{1}{x_1}, \ldots, \frac{1}{x_{br}} \right)
\tag{16}
\end{align}
Here, $x_{br}$ represents the reactance of branch $br$, and $A^\top$ is the transpose of the branch-to-bus incidence matrix, where values $\{1, -1, 0\}$ indicate the directionality and connectivity of each branch. Based on this, a value of $1$ signifies the origin (sending end) of a branch, $-1$ denotes the destination (receiving end), and $0$ indicates no direct connection between the branch and the corresponding bus. The perturbed measurement matrix after MTD is denoted as $\mathbf{H}_t'$. Each attack vector is defined as ${a} = {H}_t \mathbf{c}$ for some ${c} \in {R}^N$. The feasible attack set with bounded energy is according to $\mathcal{A} = \left\{ \mathbf{a} = \mathbf{H}_t \mathbf{c} \;\middle|\; \|\mathbf{a}\| \leq \xi_{\max},\; \mathbf{c} \in \mathbb{R}^N \right\}$. To capture the subset of attacks detectable under MTD with probability greater than a predefined threshold $\delta \in [0,1]$, it can be updated by $\mathcal{A}'(\delta) = \left\{ \mathbf{a} = \mathbf{H}_t \mathbf{c} \;\middle|\; \|\mathbf{a}\| \leq \xi_{\max},\; P_D(\mathbf{a}) > \delta,\; \mathbf{c} \in \mathbb{R}^N \right\}$. The normalized detectability ratio is then formulated via the Lebesgue measure $\lambda(\cdot)$:
\begin{equation}
\eta'(\delta) = \frac{\lambda(\mathcal{A}'(\delta))}{\lambda(\mathcal{A})}, \quad 0 \leq \eta'(\delta) \leq 1
\tag{17}
\end{equation}
This ratio quantifies the fraction of the attack space $\mathcal{A}$ that is effectively detectable post-MTD. A value of $\eta'(\delta) \approx 1$ indicates high detection coverage, while $\eta'(\delta) \approx 0$ suggests poor performance. To analyze the shift in observability space, we utilize two metrics: the Relative Subspace Distance (RSD) and the Angular Similarity Detection Indicator (ASDI). The RSD is considered to measure the average normalized deviation between the projected subspaces before and after MTD activation, and is formulated by \eqref{eq18}:
\begin{align}
\label{eq18}
\eta(\mathbf{H}_t, \mathbf{H}_t') = \frac{1}{\Lambda} \sum_{ar=1}^{\Lambda} \frac{\|\mathbf{H}_c^{ar} - \mathbf{H}_c\|}{\|\mathbf{H}_c\|}
\tag{18}
\end{align}
The ASDI is utilized to quantify the angular alignment between the most similar directions in the column spaces of $\mathbf{H}_t$ and $\mathbf{H}_t'$, and is formulated by \eqref{eq19}:
\begin{align}
\label{eq19}
\cos\left( \eta(\mathbf{H}_t, \mathbf{H}_t') \right) = \max_{\substack{\mathbf{u} \in \text{Col}(\mathbf{H}_t) \\ \mathbf{v} \in \text{Col}(\mathbf{H}_t')}} |\langle \mathbf{u}, \mathbf{v} \rangle|, \quad \|\mathbf{u}\| = \|\mathbf{v}\| = 1
\tag{19}
\end{align}
where $\Lambda$ is the number of sampled attack vectors (set to 10,000 in our simulations), and $\mathbf{H}_c^{ar}$ denotes the projection of the ${ar}^{\text{th}}$ attack vector onto the perturbed subspace. This cosine similarity metric is the primary tool used in this study to detect the presence and intensity of cyber perturbation. A low value (close to 0) signifies strong subspace divergence caused by MTD, implying improved detection. Moreover, in terms of MTD configuration considerations, two key factors are examined to optimize the MTD implementation:

\textit{1) Line Selection Strategy:} The effectiveness of MTD critically depends on which transmission branches are selected for perturbation. To assess this, the network is divided into four groups based on average loading levels. Define the average normalized loading of each line $br \in \mathcal{L}$ over the time horizon $T$ as:
\begin{align}
\label{eq20}
\phi^{'}_{br} := \frac{1}{T} \sum_{t=1}^{T} \frac{|\hat{P}_{br}^t|}{P_{br}^{\max}}
\tag{20}
\end{align}
On this basis, lines are categorized as $\mathcal{L}_\theta := \{ br \in \mathcal{L} \mid \phi^{'}_{br} \in [\theta_{\min}, \theta_{\max}]\}$
where the intervals $[\theta_{\min}, \theta_{\max}]$ correspond to (6–8\%), (8–10\%), (16–20\%), and (20–30\%). Empirical results based on Eq.~(11) indicate that perturbing lightly loaded lines leads to minimal cost increases, while perturbing heavily loaded lines significantly raises operational costs. Hence, prioritizing low-loaded branches achieves a favorable trade-off between security enhancement and economic efficiency.

\textit{2) Perturbation Magnitude (Reactance Change):} This part analyzes how the magnitude of line reactance perturbation affects both system cost and attack detectability. This can tend to a trade-off curve that compares the incremental cost of the DDM and the success rate of attack detection. The optimal configuration lies near the “knee” of this curve, where a slight increase in cost leads to a sharp gain in detectability. Mathematically, the probable increase of OPF-related cost after applying MTD is modeled as:
\begin{equation}
\label{eq21}
\text{IC-DDM}(\%) = \frac{C_{\text{DDM}} - C_0}{C_0} \times 100
\tag{21}
\end{equation}
where $C_{\text{DDM}}$ is the OPF cost with MTD and $C_0$ is the baseline cost without MTD. This analysis helps to identify the most cost-effective MTD settings.

\section{Results and Discussions}
\subsection{System Specifications}
The ML-based models are developed using the scikit-learn library \cite{kramer2016}, while DL-based models are implemented with TensorFlow \cite{abadi2016}. Additionally, all the numerical experiments and simulations in this paper are implemented using Python version 3.12, on a machine with 16 GB RAM and a 12th Gen Intel\textregistered{} Core\texttrademark{} i7-12650H 2.30 GHz processor.  
\subsection{Data Preprocessing Result}
\begin{figure}
        \centering
        \includegraphics[width=3.5in]{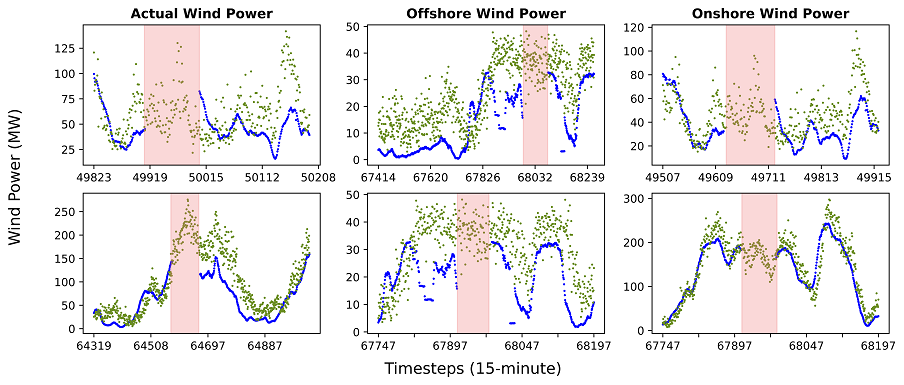}
        \caption{The prediction results of the RF regressor for three selected features.}
        \label{fig2}
\end{figure}

The results associated with the data preprocessing stage have been prepared in this subsection. During the data imputation phase, the RFR regressor outperformed other methods based on RMSE and was thus adopted for missing data reconstruction. The wind power dataset employed in this study was sourced from \cite{open2020}, encompassing multiple correlated features such as wind capacity, actual generation (segmented by offshore, onshore, and control area data). On this basis, in Fig. \ref{fig2}, the prediction results are shown for different features and timesteps of the training dataset. As shown in this figure, three selected features: Actual Wind Power, Offshore Wind Power, and Onshore Wind Power are selected to be filled by RF regressor. In this figure, each red pale area represents a period in which related data are missed. Also, blue scatter points show the actual data which are incomplete in specified time intervals. Green scatter points indicate RF predictions used to fill the gaps. The total number of samples is 201600.
Table \ref{table:regressors} shows the performance comparisons of three prevalent regressors in the data recovery stage. The RF regressor demonstrates a higher $R^2$ score, indicating its superior performance in predicting missing wind power data.
\begin{table}[H]
    \centering
    \caption{Performance comparisons of three prevalent regressors in the data recovery stage.} 
    \begin{tabular}{lcc}
        \hline
        \textbf{Regressor} & \textbf{$R^2$ Score} \\
        \hline
        RFR & 0.9504 \\
        K-nearest neighbors & 0.7694 \\
        HGBR & 0.8574 \\
        \hline
    \end{tabular}
    \label{table:regressors}
\end{table}
\subsection{Att-BiLSTM Prediction Outcomes}
\begin{figure}[h!]
\centering
\includegraphics[width=0.85\linewidth]{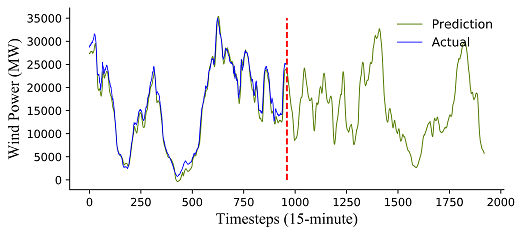}
\caption{Wind power prediction used by the proposed modified Att-BiLSTM model for ten days ahead.}
\label{fig3}
\end{figure}

Following the preprocessing phase, missing values in the wind power dataset were imputed using a regression-based approach to ensure data consistency. The refined dataset was subsequently utilized to train various wind power prediction models. As discussed earlier, four advanced deep learning architectures—each enhanced through OPTUNA-based hyperparameter tuning—are employed and evaluated on the same dataset. In Fig. \ref{fig3}, actual and Att-BiLSTM predicted values for wind power are observable, and this figure depicts the wind power forecasting for ten days attained by the modified Att-BiLSTM model. 

Four different criteria are employed to assess the performance and delineate the most preferred architecture, gathered in Table \ref{tab:performance_comparison}. As it is clear and expected, the proposed modified Att-BiLSTM records much better in comparison with others.
\begin{table}[!t]
\centering
\caption{Performance measures comparisons between the proposed model and some other prevalent methods.}
\begin{tabularx}{\linewidth}{lXXXX}
\toprule
\textbf{Evaluation metrics} & \textbf{Att-BiLSTM} & \textbf{LSTM \cite{LSTM}} & \textbf{RNN \cite{narayanan2024}} & \textbf{GRU \cite{zhang2024(2)}} \\
\midrule
$R^2$               & 0.9878 & 0.9808 & 0.9649 & 0.9719 \\
MAPE                & 0.1005 & 0.1246 & 0.1559 & 0.1930 \\
NMAE                & 1.4284 & 1.7288 & 2.3359 & 2.0894 \\
NRMSE               & 1.9385 & 2.4351 & 3.2992 & 2.9511 \\
\bottomrule
\end{tabularx}
\label{tab:performance_comparison}
\end{table}
\subsection{GAN Model Results} 
\begin{figure}[h!]
\centering
\includegraphics[width=0.45\textwidth]{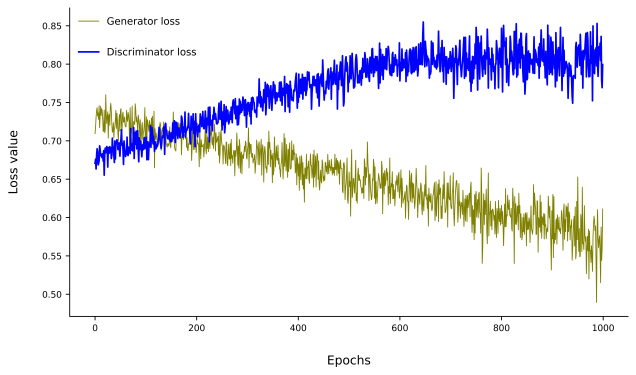}
\caption{Fitness process of the proposed ConvGAN model including generator and discriminator during 1000 epochs.}
\label{fig4}
\end{figure}

\begin{figure}[h!]
\centering
\includegraphics[width=0.5\textwidth]{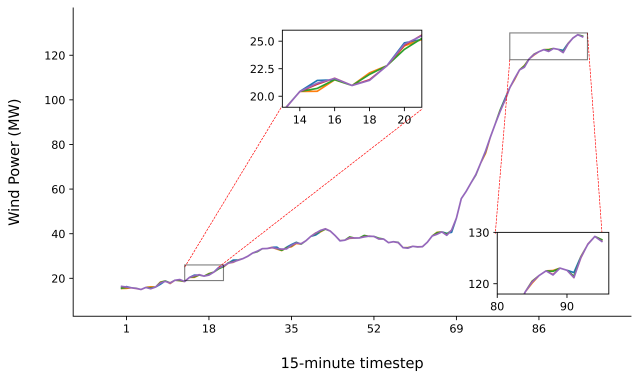} 
\caption{1000 randomly generated scenarios of wind power using the ConvGAN model.}
\label{fig5}
\end{figure}

When the wind power prediction stage is successfully completed by the selected preferable architecture, the proposed ConvGAN model is responsible for estimating wind power fluctuations and uncertainty modeling. As illustrated in Fig. \ref{fig4}, this deep-based generation model must be trained until the generator component can defeat its rival, the discriminator. After training ConvGAN model, it would be able to generate realistic synthetic scenarios. Accordingly, Fig. \ref{fig5} shows 1000 scenarios generated by the proposed ConvGAN model for a day of wind power prediction. 

\begin{figure*}[!t] % 'h' here ensures that the figure will approximately be located here in the text.
\centering
\includegraphics[width=0.85\linewidth]{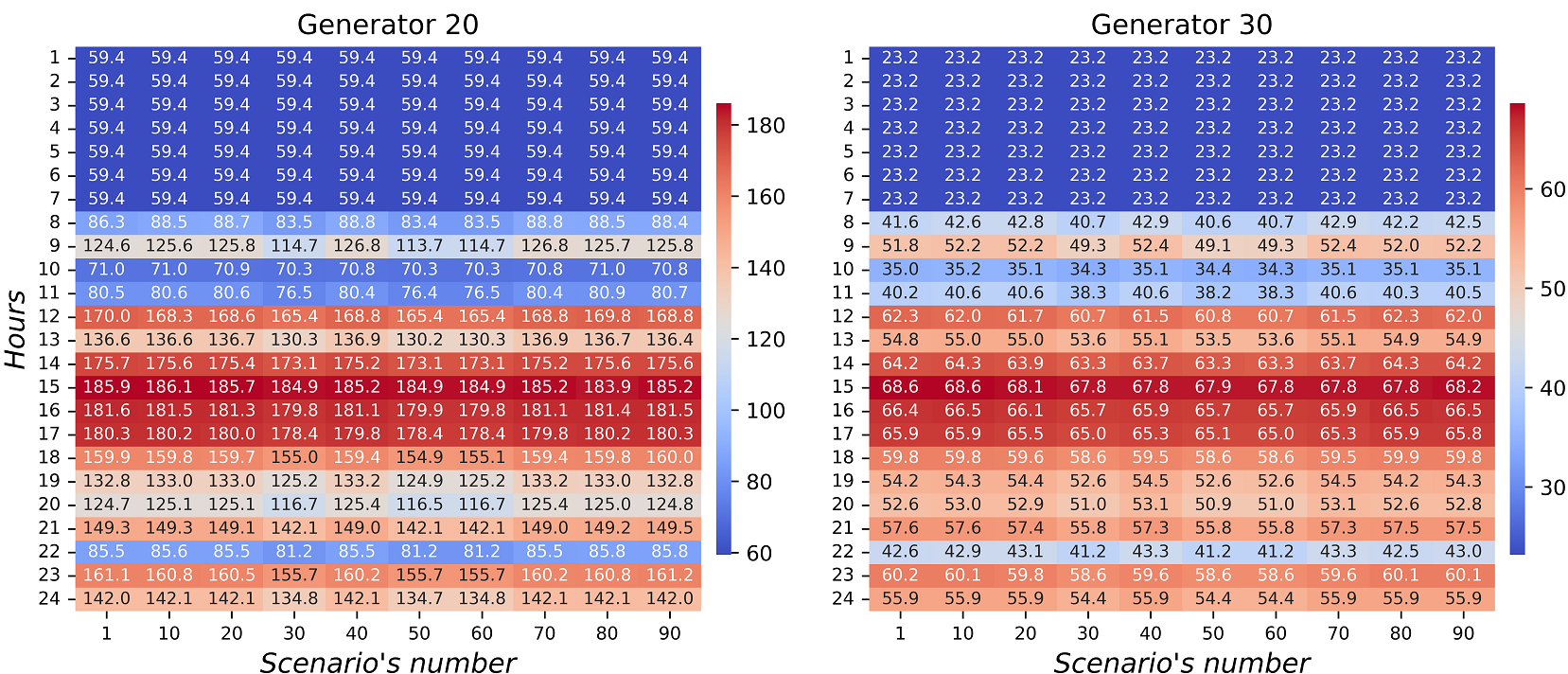}
\caption{Cost-benefit optimal scheduling of two sample generators via ten sample scenarios in Case1.}
\label{fig6}
\end{figure*}

\begin{figure*}[!t]
\centering
\includegraphics[width=0.85\linewidth]{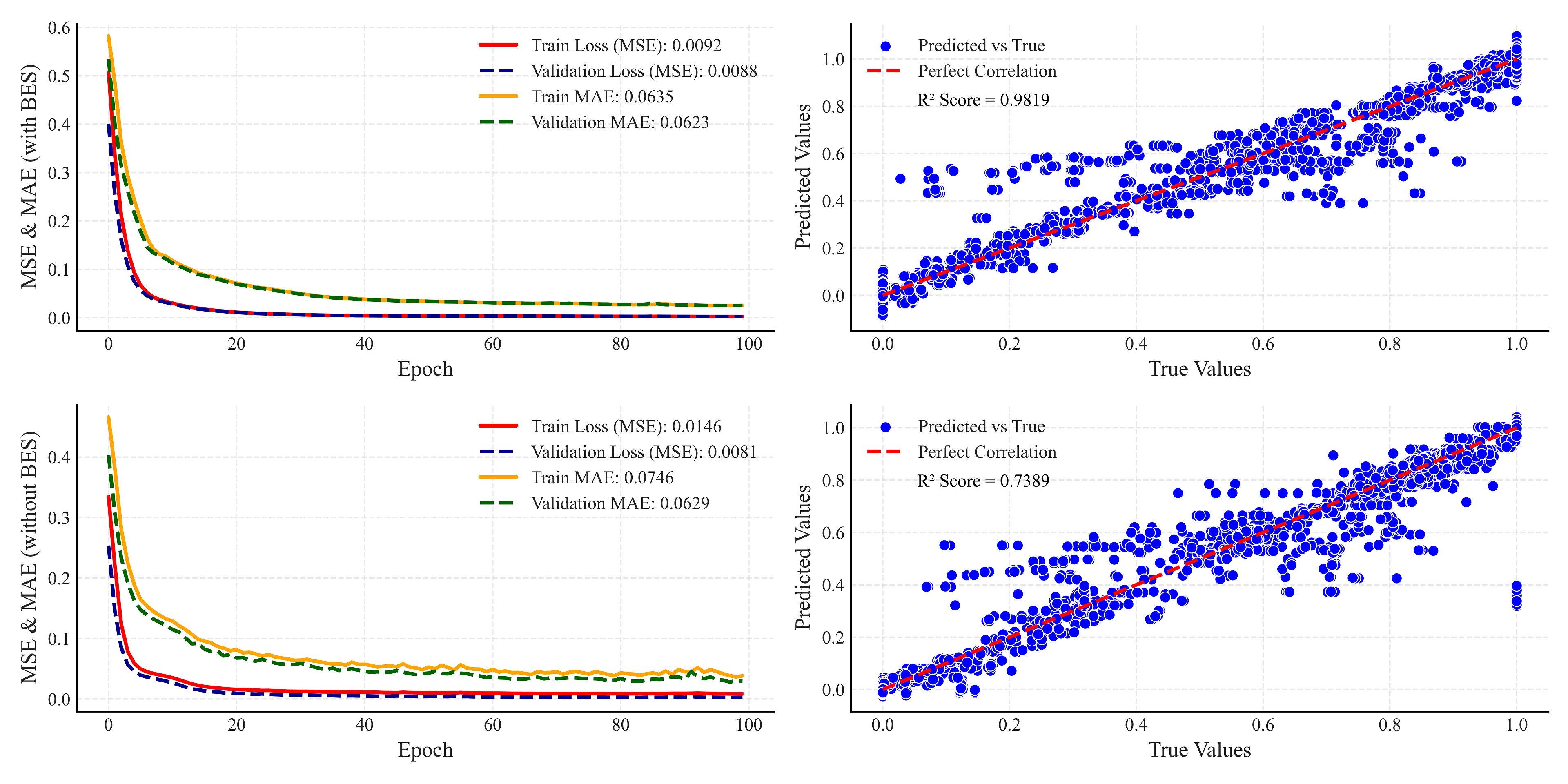}
\caption{Fitness evaluation of proposed deep-based BiLSTM-CNN model along with correlation for both Case1 and Case2.}
\label{fig7}
\end{figure*}
\subsection{Predictive Optimal Scheduling Model Results} 
\begin{figure}[h!]
\centering
\includegraphics[width=0.85\linewidth]{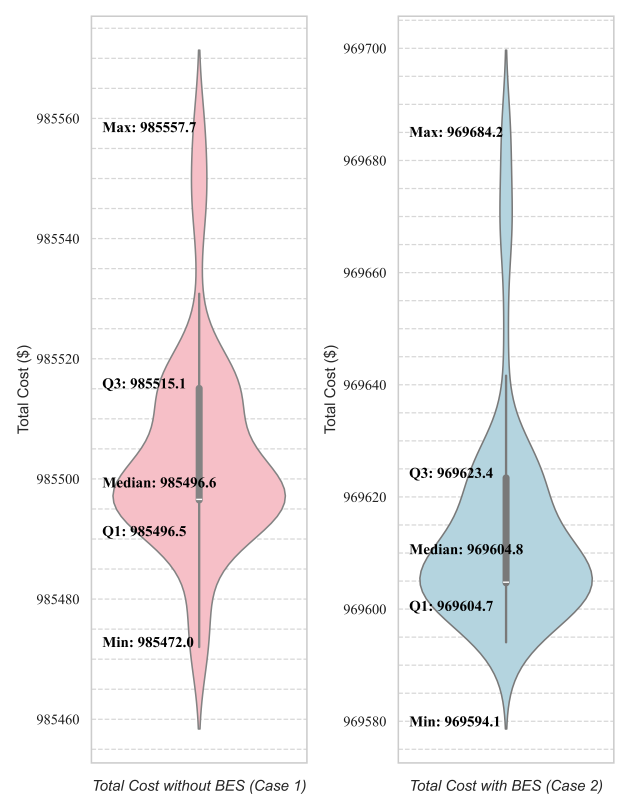}
\caption{Fitness evaluation of cost-benefit optimization along with correlation for both Case 1 and Case 2.}
\label{fig8}
\end{figure}

In Fig. \ref{fig6}, the heatmaps illustrate the output scheduling across 10 scenarios for two generators over a 24-hour period. Each color transition from blue to red represents an increase in output level, with blue indicating low output during night hours and red representing peak output hours. The figure highlights the varying operational demands met by each generator, particularly during peak hours (approximately hours 10-19).  
This visualization confirms the model's effectiveness in adjusting generator output across scenarios to meet system demands, ensuring economic efficiency and reliability. Fig. \ref{fig7} illustrates the time-series predictions and correlations for the two cases, achieved using a deep-based Att-BiLSTM-CNN architecture. As expected, the incorporation of BES leads to reduced operating costs by enhancing wind power integration. 
To illustrate this, statistical parameters are presented using box plots, along with their frequency distribution depicted as a violin plot in Fig. \ref{fig8}. The violin plot highlights the density of the cost values. In both cases, the majority of the costs cluster around their respective medians, with Case 2 showing a more concentrated and narrow distribution due to the stabilizing effect of BES.
Given the objective function results in Table \ref{tab:11}, along with wind curtailment changes across different scenarios for varying BES sizes and charging/discharging potentials, it can be concluded that investing in larger BES capacities leads to both lower costs and reduced losses. 
\begin{table}[!t]
\caption{THE IMPACTS OF BES SIZE ON THE TOTAL COST AND THE CURTAILED WIND POWER.}
\centering
\renewcommand{\arraystretch}{1} % Adjust row spacing for better readability
\begin{tabularx}{\linewidth}{Xcc}
\toprule
\textbf{BES Capacity (MWh)} & \textbf{Total Cost (\$)} & \textbf{Wind Curtailment (MW)} \\
\midrule
500  & 969,652.28 & 102.89 \\
1000 & 955,379.12 & 94.27 \\
1500 & 948,197.74 & 0 \\
\bottomrule
\end{tabularx}
\label{tab:11}
\end{table}
\subsection{Cybersecurity-based Optimization Results} 
\begin{figure*}[!t]
\centering
\includegraphics[width=0.75\linewidth]{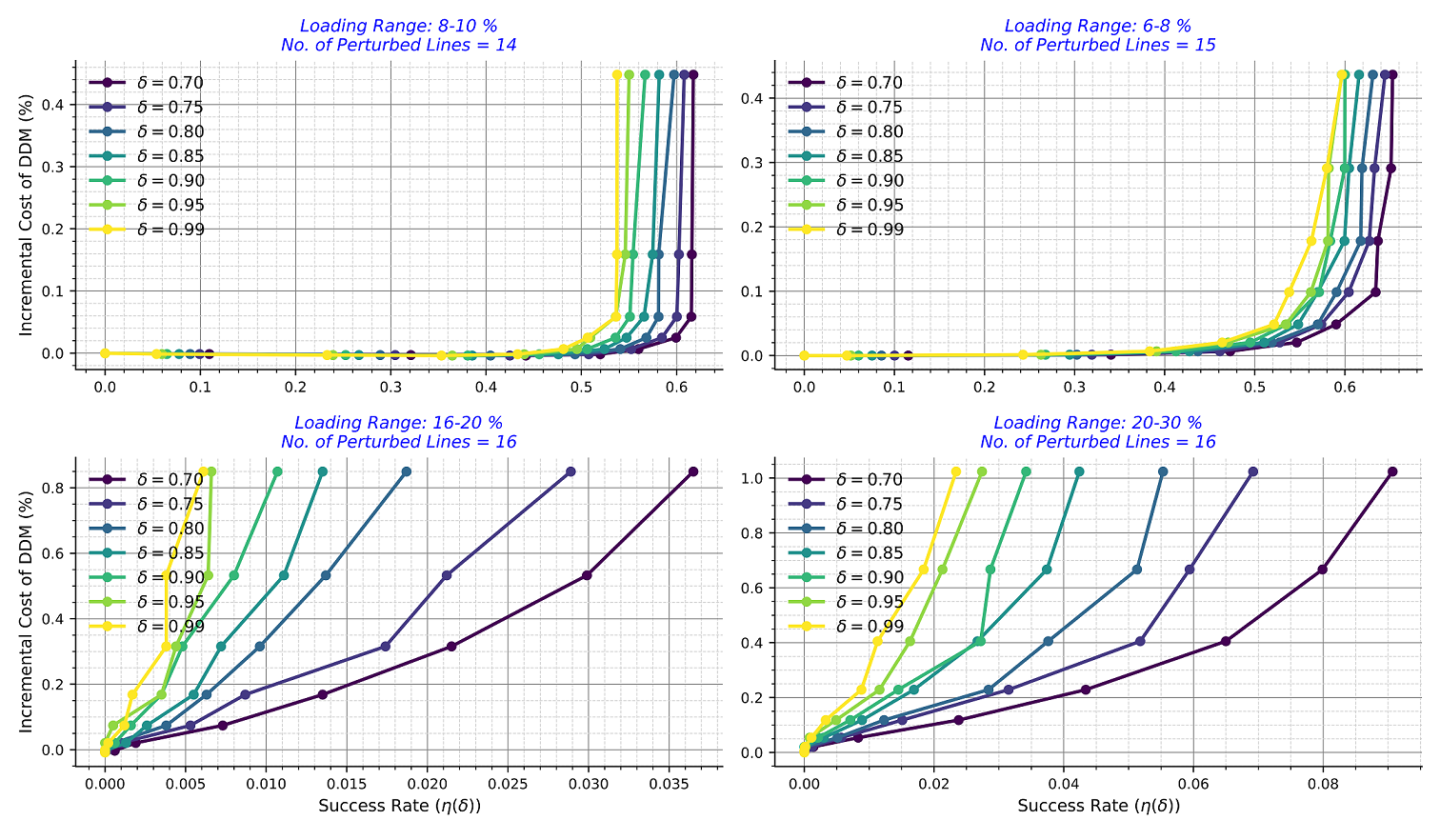}
\caption{Trade-off analysis between success rate and incremental cost of DDM across four different loading ranges without using storage.}
\label{fig9}
\end{figure*}

The results related to the cybersecurity-based optimization is provided in this subsection. Fig. \ref{fig9} visualizes how the success rate ($\eta(\delta)$) impacts the incremental cost of DDM (\%) for different values of $\delta$, ranging from 0.7 to 0.99. More conservative security filtering (higher $\delta$ values) results in a lower ability to reach peak effectiveness. Apart from $\delta$ values, the selection of lines and their loadings directly affects effectiveness. In addition to the impacts of the branch network on the objective function, the engagement of BES can also influence the outcomes, which should be considered. Consequently, two scenarios—utilizing/not utilizing BES—are conducted, with their detailed results presented in Table \ref{P11}. Although BES utilization does not significantly reduce total costs compared to the economic scheduling mode, in most instances (15 instances), it results in a lower cost increment when compared to scenarios without BES utilization.

\begin{table*}[htbp]
\centering
\caption{OPTIMAL COST INCREMENT AND SUCCESS RATE FOR DIFFERENT CASES WITH THEIR CORRESPONDING $\delta$ VALUES, DIFFERENTIATED BY CONFIGURATIONS WITH/WITHOUT BES.}
\begin{tabular}{@{}cccccccccccc@{}}
\hline
\multicolumn{5}{c}{\textbf{Loading Range: 8-10\%}} & & \multicolumn{5}{c}{\textbf{Loading Range: 6-8\%}} & \\
\cmidrule{1-5} \cmidrule{7-11}
\multicolumn{2}{c}{Optimal Success Rate} & \multicolumn{2}{c}{Optimal Cost Increment} & $\delta$ & & 
\multicolumn{2}{c}{Optimal Success Rate} & \multicolumn{2}{c}{Optimal Cost Increment} & $\delta$ &  \\
No BES & BES & No BES & BES &  &  & No BES & BES & No BES & BES &  \\
\hline
0.5326 & 0.3874 & 0.0299 & 0.0242 & 0.70 &  & 0.4053 & 0.3859 & 0.0650 & 0.0670 & 0.70 &  \\
0.3155 & 0.6276 & 0.0174 & 0.0212 & 0.75 &  & 0.4053 & 0.3859 & 0.0518 & 0.0499 & 0.75 &  \\
0.3155 & 0.3874 & 0.0096 & 0.0121 & 0.80 &  & 0.6670 & 0.3859 & 0.0513 & 0.0346 & 0.80 &  \\
0.5326 & 0.3874 & 0.0111 & 0.0099 & 0.85 &  & 0.6670 & 0.6321 & 0.0374 & 0.0377 & 0.85 &  \\
0.5326 & 0.6276 & 0.0080 & 0.0094 & 0.90 &  & 0.4053 & 0.6321 & 0.0272 & 0.0295 & 0.90 &  \\
0.5326 & 0.2172 & 0.0064 & 0.0031 & 0.95 &  & 0.4053 & 0.3859 & 0.0163 & 0.0183 & 0.95 &  \\
0.3155 & 0.6276 & 0.0038 & 0.0044 & 0.99 &  & 0.2286 & 0.3859 & 0.0088 & 0.0137 & 0.99 &  \\
\hline
\multicolumn{5}{c}{\textbf{Loading Range: 16-20\%}} & & \multicolumn{5}{c}{\textbf{Loading Range: 20-30\%}} & \\
\cmidrule{1-5} \cmidrule{7-11}
\multicolumn{2}{c}{Optimal Success Rate} & \multicolumn{2}{c}{Optimal Cost Increment} & $\delta$ & &
\multicolumn{2}{c}{Optimal Success Rate} & \multicolumn{2}{c}{Optimal Cost Increment} & $\delta$ &  \\
No BES & BES & No BES & BES &  &  & No BES & BES & No BES & BES &  \\
\hline
0.0985 & 0.1878 & 0.6341 & 0.6458 & 0.70 &  & 0.0585 & 0.0423 & 0.6154 & 0.6099 & 0.70 &  \\
0.1780 & 0.1026 & 0.6270 & 0.6075 & 0.75 &  & 0.0585 & 0.0423 & 0.6002 & 0.5938 & 0.75 &  \\
0.1780 & 0.1878 & 0.6175 & 0.6132 & 0.80 &  & 0.0585 & 0.0423 & 0.5810 & 0.5795 & 0.80 &  \\
0.1780 & 0.1878 & 0.5995 & 0.5984 & 0.85 &  & 0.0585 & 0.1472 & 0.5658 & 0.5772 & 0.85 &  \\
0.0985 & 0.1878 & 0.5713 & 0.5857 & 0.90 &  & 0.0585 & 0.0423 & 0.5509 & 0.5461 & 0.90 &  \\
0.1780 & 0.1878 & 0.5812 & 0.5697 & 0.95 &  & 0.0585 & 0.0423 & 0.5361 & 0.5436 & 0.95 &  \\
0.1780 & 0.1878 & 0.5628 & 0.5574 & 0.99 &  & 0.0585 & 0.0423 & 0.5364 & 0.5311 & 0.99 &  \\
\hline
\end{tabular}
\label{P11}
\end{table*}
\section{Conclusion}
This paper presents a predictive optimal scheduling framework aimed at maintaining the economic and reliable operation of power systems through the integration of Battery Energy Storage (BES) and a robust cybersecurity mechanism. The main contributions of the proposed model include: (i) a BiLSTM-based short-term wind forecasting module with high accuracy, (ii) a ConvGAN-driven scenario generation for capturing wind uncertainty, (iii) a two-layered optimization model addressing both economic and cybersecurity goals, and (iv) the use of BES to support both economic scheduling and cybersecurity through the Moving Target Defense (MTD) strategy. 
The proposed predictive model achieved impressive accuracy, with 98.78\% precision in wind forecasting and final predictive model’s accuracies of 98\% and 74\% with and without BES integration, respectively. 
The results demonstrate that incorporating BES into the power system can obtain lower total costs by at least 1.6\% (with minimum storage capacity) and up to 3.78\% (with maximum storage capacity). Moreover, wind curtailment was reduced by 5.54\% at the lowest and entirely eliminated (100\%) at the highest level of utilized BES capacities. Additionally, as detailed in the sensitivity analysis, the impact of three key parameters is explored—which are wind prices and BES charging/discharging costs—on the objective function, providing valuable insights. On this basis, As BES operational costs increase, the economic efficiency of BES diminishes, making it less attractive for large-scale deployment. Additionally, it is revealed that BES utilization can mitigate the cost increment related to the DDM implementation as well. Furthermore, the proposed MTD model's cost-security tradeoffs offer valuable insights for optimal decision-making based on varying detectability thresholds. Specifically, we address two key questions: how and where to apply perturbations to line reactances using angular similarity detection and the knee method.

\bibliographystyle{IEEEtran}
\bibliography{Reference}

\end{document}